\documentclass[12pt]{article}

\begin{document}

\title{Izergin-Korepin determinant reloaded}
\author{Yu.~G.~Stroganov\\
\small \it Institute for High Energy Physics\\[-.5em]
\small \it 142284 Protvino, Moscow region, Russia}
\date{}

\maketitle

\begin{abstract}
We consider the Izergin-Korepin determinant~\cite{iz1} together with
another determinant which was invented by Kuperberg~\cite{Ku2}.
He used these determinants to prove a formula for the total number
of  half-turn symmetric alternating sign matrices of even order
conjectured by Robbins~\cite{Rob}.
By developing further the method that was described in our previous paper~\cite{my1},
we obtain a closed nonlinear recurrence system for these determinants.
It can be used in various ways. For example, in this paper, we obtain 
formula (\ref{Result}) for the refined enumeration of  half-turn symmetric
alternating sign matrices of even order.[

\end{abstract}

\newpage 

\begin{center}{\bf1. Six-vertex model with domain wall boundary and Izergin-Korepin
determinant}
\end{center}

Let us consider the inhomogeneous six-vertex model on a square lattice.
The states of the model are given by assigning arrows to each edge of an 
$n \times n$ square lattice so that at each vertex, two arrows go in and two go out
(\it ice condition \rm).
Spectral parameters $\{x_1,x_2,...,x_{n}\}$ and $\{y_1,y_2,...,y_{n}\}$ are attached
to the horizontal and vertical lines respectively.
Edges point inward at the sides and outward at the top and bottom as in figure 1.
With these boundary conditions which are called \it domain wall  boundary conditions
\rm \cite{kor1} the states of the model are in bijection with a certain set of
matrices, called \it alternating sign matrices \rm
.

\vspace{0.5cm}

\begin{picture}(0,200)
\put(40,60){$x_n$}
\put(40,150){$x_2$}
\put(40,180){$x_1$}

\put(60,60){\vector(1,0){18}}
\put(78,60){\line(1,0){12}}

\put(60,150){\vector(1,0){18}}
\put(78,150){\line(1,0){12}}

\put(60,180){\vector(1,0){18}}
\put(78,180){\line(1,0){12}}

\put(90,60){\line(1,0){60}}
\put(90,150){\line(1,0){60}}
\put(90,180){\line(1,0){60}}

\put(160,60){\dots}
\put(160,150){\dots}
\put(160,180){\dots}

\put(180,60){\line(1,0){30}}
\put(180,150){\line(1,0){30}}
\put(180,180){\line(1,0){30}}

\put(240,60){\vector(-1,0){18}}
\put(222,60){\line(-1,0){12}}

\put(240,150){\vector(-1,0){18}}
\put(222,150){\line(-1,0){12}}

\put(240,180){\vector(-1,0){18}}
\put(222,180){\line(-1,0){12}}

\put(88,20){$y_1$}
\put(118,20){$y_2$}
\put(208,20){$y_{n}$}

\put(90,60){\vector(0,-1){18}}
\put(90,42){\line(0,-1){12}}

\put(120,60){\vector(0,-1){18}}
\put(120,42){\line(0,-1){12}}

\put(210,60){\vector(0,-1){18}}
\put(210,42){\line(0,-1){12}}

\put(90,60){\line(0,1){30}}
\put(120,60){\line(0,1){30}}
\put(210,60){\line(0,1){30}}

\put(89,100){\vdots}
\put(119,100){\vdots}
\put(209,100){\vdots}

\put(90,120){\line(0,1){60}}
\put(120,120){\line(0,1){60}}
\put(210,120){\line(0,1){60}}

\put(90,180){\vector(0,1){18}}
\put(90,198){\line(0,1){12}}

\put(120,180){\vector(0,1){18}}
\put(120,198){\line(0,1){12}}

\put(210,180){\vector(0,1){18}}
\put(210,198){\line(0,1){12}}

\put(160,100){$\ddots$}
\put(120,0){Figure 1} 

\end{picture}

\vspace{0.5cm}

There are six possible configurations of the arrows on the edges for a given vertex as
in figure 2. 

\vspace{0.5cm}

\begin{picture}(0,60)
\put(0,40){\vector(1,0){13}}
\put(13,40){\line(1,0){7}}
\put(20,40){\vector(1,0){13}}
\put(33,40){\line(1,0){7}}
\put(20,20){\vector(0,1){13}}
\put(20,33){\line(0,1){7}}
\put(20,40){\vector(0,1){13}}
\put(20,53){\line(0,1){7}}
\put(43,38){\tiny =}
\put(70,40){\vector(-1,0){13}}
\put(57,40){\line(-1,0){7}}
\put(90,40){\vector(-1,0){13}}
\put(77,40){\line(-1,0){7}}
\put(70,60){\vector(0,-1){13}}
\put(70,47){\line(0,-1){7}}
\put(70,40){\vector(0,-1){13}}
\put(70,27){\line(0,-1){7}}
\put(93,38){\tiny = \small a}
\put(120,40){\vector(1,0){13}}
\put(133,40){\line(1,0){7}}
\put(140,40){\vector(1,0){13}}
\put(153,40){\line(1,0){7}}
\put(140,40){\vector(0,-1){13}}
\put(140,27){\line(0,-1){7}}
\put(140,60){\vector(0,-1){13}}
\put(140,47){\line(0,-1){7}}
\put(163,38){\tiny =}
\put(190,40){\vector(-1,0){13}}
\put(177,40){\line(-1,0){7}}
\put(210,40){\vector(-1,0){13}}
\put(197,40){\line(-1,0){7}}
\put(190,40){\vector(0,1){13}}
\put(190,53){\line(0,1){7}}
\put(190,20){\vector(0,1){13}}
\put(190,33){\line(0,1){7}}
\put(213,38){\tiny = \small b}
\put(240,40){\vector(1,0){13}}
\put(253,40){\line(1,0){7}}
\put(280,40){\vector(-1,0){13}}
\put(267,40){\line(-1,0){7}}
\put(260,40){\vector(0,-1){13}}
\put(260,27){\line(0,-1){7}}
\put(260,40){\vector(0,1){13}}
\put(260,53){\line(0,1){7}}
\put(283,38){\tiny =}
\put(310,40){\vector(-1,0){13}}
\put(297,40){\line(-1,0){7}}
\put(310,40){\vector(1,0){13}}
\put(323,40){\line(1,0){7}}
\put(310,60){\vector(0,-1){13}}
\put(310,47){\line(0,-1){7}}
\put(310,20){\vector(0,1){13}}
\put(310,33){\line(0,1){7}}
\put(333,38){\tiny = \small c}
\put(160,5){Figure 2}
\end{picture}

\vspace{0.5cm}
  
The Boltzmann weights which are assigned to every vertex of the lattice have the form:
\begin{eqnarray}
\label{weights}
&&a(x-y)=\sin (\eta/2+x-y), \quad  b(x-y)=\sin(\eta/2-x+y),\\
&&c(x-y)=\sin \eta. \nonumber 
\end{eqnarray}
The different letters in Figure 1 correspond to the different functions introduced
in equation (\ref{weights}). The $x$ and $y$ are the horizontal and vertical spectral
parameters, respectively, which depend on the vertex position. 
The $\eta$, sometimes called \it crossing parameter \rm has the same value for all
vertices.

The weight of a state of the model is the product of the weights of its vertices,
and the \it partition function (state sum) \rm is the total weight of all states.

Let $Z_n(x_1,x_2,...,x_{n}; y_1,y_2,...,y_{n})$ be the resulting state sum.  
Izergin \cite{iz1} (using the above mentioned work of Korepin 
\cite{kor1} ) found a determinant representation for Z.
\begin{eqnarray}
\label{IKdet}
&&Z_n(\{x\},\{y\})=\prod_{1 \le i,j \le {n}} \sin(\eta/2+x_i-y_j) \>
\prod_{1 \le i,j \le {n}} \sin(\eta/2-x_i+y_j) \times \nonumber \\
&& \times \frac{ \mbox{det} M(\{x\},\{y\})} 
{\prod_{1 \le i < i^{\prime} \le {n}} \sin(x_i-x_{i^{\prime}})
\prod_{1 \le j < j^{\prime} \le {n}} \sin(y_{j^{\prime}}-y_j)},
\end{eqnarray}
where the entries of an $n \times n$  matrix $M(\{x\},\{y\})$ are
\begin{eqnarray}
&&M_{i,j}=\frac{\sin \eta}{\sin(\eta/2+x_i-y_j) \> \sin(\eta/2-x_i+y_j)},
\quad i,j=1,2,...,n. \nonumber
\end{eqnarray}

Let us write
\begin{eqnarray}
\label{xyunion}
&&u_0 \equiv u\>\> \mbox{for}\>\> x_1,\nonumber \\
&& u_i\>\> \mbox{for}\>\>  x_{i+1}, \quad i=1,...,n-1, \\
&&\mbox{and} \>\>u_{i+n-1}\>\>\mbox{for}\>\> y_i, \quad  i=1,...,n. \nonumber
\end{eqnarray}

In our recent paper \cite{my1} we considered the case $\eta= 2\pi/3$ and introduced
the sequence of functions $f_n$ defined by
\begin{eqnarray}
 \label{fdef}
&&f_n(u) = Z_n(u)\> \prod_{i=1}^{2n-1} \sin (u-u_i),
\end{eqnarray}
where, for brevity, we write $f_n(u)$ and $Z_n(u)$ rather than $f_n(u,u_1,...u_{2n-1})$
and $Z_n(u,u_1,...u_{2n-1})$ suppressing the variables $u_i,\>\>i=1,...,2n-1$. 

By using some properties of the state sum $Z_n(u)$ \cite{kor1,KIB} we found 
that the function $f_n(u)$ is trigonometric polynomial of degree $3n-2$
which can be expressed by the finite Fourier sum:
\begin{eqnarray}
 \label{fFourier}
&&f_n(u) = \sum_{k=1,k \ne 3\kappa}^{3n-1} a_k \exp{i(3n-2k)u}.
\end{eqnarray}

We showed also that equations (\ref{fdef}) and (\ref{fFourier}) fix the function $f_n(u)$, and 
consequently, the state sum $Z_n(u)$ up to an arbitrary constant multiplier and 
calculated the latter for several cases related to enumeration of alternating sign 
matrices (see also the subsequent papers \cite{my2,RSasm}).

In the  same paper \cite{my1} we found that in the case of $\eta= 2\pi/3$ the state sum
$Z_n(u_0\equiv u,u_1,u_2,...,u_{2n-1})$ is symmetric in variables $\{u_0 \equiv u,u_1,u_2,...,u_{2n-1}\}$
and satisfy another determinant representation
\begin{eqnarray}
 \label{Zdet}
&&Z_n(\{u\}) =  z_n \biggl (\prod_{0 \le j<j^{\prime} \le 2n-1} 
\sin (u_j-u_{j^{\prime}})\biggr )^{-1} P,
\end{eqnarray}
where
\begin{eqnarray}
&&P= \mbox{det} \>\>\begin{array}{|ccccc|}
t_0^{3n-2}&t_1^{3n-2}&t_2^{3n-2}&\dots&t_{2n-1}^{3n-2} \\
t_0^{3n-4}&t_1^{3n-4}&t_2^{3n-4}&\dots&t_{2n-1}^{3n-4}  \\
t_0^{3n-8}&t_1^{3n-8}&t_2^{3n-8}&\dots&t_{2n-1}^{3n-8}  \\
\vdots&\vdots&\vdots&\ddots&\vdots   \\
t_0^{2-3n}&t_1^{2-3n}&t_2^{2-3n}&\dots&t_{2n-1}^{2-3n}
\end{array},  \nonumber
\end{eqnarray}
$t_j=\exp (i\>u_j),\>j=0,1,...,2n-1$ and $z_n$ does not depend on $\{u\}$
(see also \cite{Okada} and references therein).

In this paper we continue to consider the case $\eta= 2\pi/3$ and find the unlinear 
recurrence for the state sum $Z_n(u)$.

\begin{center}{\bf2. Companion of Izergin-Korepin construction}
\end{center}
We consider together with the state sum $Z_n(\{x\},\{y\})$ a similar function
$V_n(\{x\},\{y\})$ related to $n \times n$ matrix $N$:

\begin{eqnarray}
\label{Kupdet}
&&V_n(\{x\},\{y\})=\prod_{1 \le i,j \le {n}} \sin(\eta/2+x_i-y_j) \>
\prod_{1 \le i,j \le {n}} \sin(\eta/2-x_i+y_j) \times \nonumber \\
&& \times \frac{\mbox{det} N(\{x\},\{y\})} 
{\prod_{1 \le i < i^{\prime} \le {n}} \sin(x_i-x_{i^{\prime}})
\prod_{1 \le j < j^{\prime} \le {n}} \sin(y_{j^{\prime}}-y_j)},
\end{eqnarray}
where the entries of  matrix $N(\{x\},\{y\})$ are
\begin{eqnarray}
\label{Nentries}
&&N_{i,j}=\frac{1}{\sin(\eta/2+x_i-y_j)}+\frac{1}{\sin(\eta/2-x_i+y_j)},\quad i,j=1,2,...,n.
\end{eqnarray}

The matrix $N$ is proportional to Kuperberg's matrix $M_{HT}^+$.  
The latter enters the expression (see Theorem 10 of paper \cite{Ku2}) for the partition
function $Z_n^{HT}(\{x\},\{y\})$ of square ice model with special boundary conditions
(see Figure 7 of paper \cite{Ku2}). This partition function gives the weighted enumeration of 
the half-turn symmetric alternating sign matrices.
Robbins conjectured the 1-enumeration and established the 2-enumeration of this class of
matrices~\cite{Rob}. In the case of even order matrices Kuperberg established the determinant
formula which is, by the way, the product of equations (\ref{IKdet}) and (\ref{Kupdet}).
He used it, among other things, to obtain the 1-,2-, and 3-enumerations of
the half-turn symmetric alternating sign matrices.

According to Theorem 10 of paper~\cite{Ku2}, the function $V_n(\{x\},\{y\})$ is equal
to the ratio of the partition function
$Z_n^{HT}(\{x\},\{y\})$ to the partition function $Z_n(\{x\},\{y\})$.
Using the method described in the paper~\cite{my1} we find that the new functions 
 $g_n(u)$ defined by
\begin{eqnarray}
 \label{gdef}
&&g_n(u) = V_n(u)\> \prod_{i=1}^{2n-1} \sin (u-u_i),
\end{eqnarray}
can be expressed by the finite Fourier sum
\begin{eqnarray}
 \label{gFourier}
&&g_n(u) = \sum_{k=1,k \ne 3\kappa+2}^{3n} b_k \exp{i(3n-2k+1)u}.
\end{eqnarray}
Compare this equations with equations (\ref{fdef}) and (\ref{fFourier}).
We write, for brevity, $g_n(u)$ and $V_n(u)$ rather than $g_n(u,u_1,...u_{2n-1})$
and $V_n(u,u_1,...u_{2n-1})$ suppressing the variables $u_i,\>\>i=1,...,2n-1$ which
are defined by equation (\ref{xyunion}).

Using the technique of the paper~\cite{my1} one can show that 
equations (\ref{gdef}) and (\ref{gFourier}) fix the function $g_n(u)$, and 
consequently, the function  $V_n(u)$ up to an arbitrary constant multiplier.
The latter  is symmetric in variables $\{u_0 \equiv u,u_1,u_2,...,u_{2n-1}\}$
and satisfies the  determinant representation similar to Izergin-Korepin one:
\begin{eqnarray}
 \label{Vdet}
&&V_n(\{u\}) =  v_n \biggl (\prod_{0 \le j<j^{\prime} \le 2n-1} 
\sin (u_j-u_{j^{\prime}})\biggr )^{-1} R,
\end{eqnarray}
where
\begin{eqnarray}
&&R= \mbox{det} \>\>\begin{array}{|ccccc|}
t_0^{3n-1}&t_1^{3n-1}&t_2^{3n-1}&\dots&t_{2n-1}^{3n-1} \\
t_0^{3n-5}&t_1^{3n-5}&t_2^{3n-5}&\dots&t_{2n-1}^{3n-5}  \\
t_0^{3n-7}&t_1^{3n-7}&t_2^{3n-7}&\dots&t_{2n-1}^{3n-7}  \\
\vdots&\vdots&\vdots&\ddots&\vdots   \\
t_0^{1-3n}&t_1^{1-3n}&t_2^{1-3n}&\dots&t_{2n-1}^{1-3n}
\end{array}, \nonumber
\end{eqnarray}
$t_j=\exp (i\>u_j),\>j=0,1,...,2n-1$ and $v_n$ does not depend on $\{u\}$.

\begin{center}{\bf3. The main idea of the paper}
\end{center}

The finite Fourier sums (\ref{fFourier}) and (\ref{gFourier}) for $f_n(u)$ and $g_n(u)$ correspondingly
are a complement one of another.
By using them together we can obtain the recurrence in $n$ for the state sum $Z_n(u)$ and
its companion $V_n(u)$.

For the illustration let us consider the case $n=2$:
\begin{eqnarray}
&&f_2(u) = a_1 e^{4iu} + a_2 e^{2iu}+ a_4 e^{-2iu}+ a_5 e^{-4iu}, \nonumber \\
&&g_2(u) = b_1 e^{5iu} + b_3 e^{iu}+ b_4 e^{-iu}+ b_6 e^{-5iu}. \nonumber 
\end{eqnarray}
Recall (see equations (\ref{fdef}) and (\ref{gdef})) that the product 
$$\sin(u-u_1) \sin(u-u_2) \sin(u-u_3)$$ 
divides both of these polynomials.
This circumstance allows us to obtain the coefficients $a_i$ and $b_i$ up to an arbitrary constant
multiplier.
We can, for example, to solve two homogeneous systems of linear equation:
\begin{eqnarray}
&&f_2(u_i) = 0,\quad i=1,2,3, \nonumber \\
&&g_2(u_i) = 0,\quad i=1,2,3. \nonumber
\end{eqnarray}
This way leads us to determinants $P$ and $R$ (see equations (\ref{Zdet}) and (\ref{Vdet})
correspondingly).

Let us proceed in a different way.
Let us use the preceding pair
\begin{eqnarray}
&&f_1(u) = \tilde{a}_1 e^{iu} + \tilde{a}_2 e^{-iu}, \nonumber \\
&&g_1(u) = \tilde{b}_1 e^{2iu} + \tilde{b}_3 e^{-2iu}. \nonumber 
\end{eqnarray}
We see first that the combination
\begin{eqnarray}
 \label{combination}
&&X g_1(u)+(Y e^{3iu}+\bar{Y} e^{-3iu})f_1(u),
\end{eqnarray}
where $X,Y$ and $\bar{Y}$ do not depend on $u$ has the same Fourier structure as $f_2(u)$.
Second, the monomial   $\sin(u-u_1)$ divides both of functions $f_1(u)$ and $g_1(u)$, and 
consequently divides the whole expression (\ref{combination}).
By choosing ratios $X,Y$ and $\bar{Y}$ so that the remaining factor 
$$\sin(u-u_2) \sin(u-u_3)$$ also divides the expression (\ref{combination}) we obtain the function
which has to be $f_2(u)$ up to an arbitrary constant multiplier.
Generalizing this idea to the arbitrary $n$ case we easily obtain the pair of nonlinear 
recurrence relation
\begin{eqnarray}
 \label{fgrec}
&&f_{n+1}({\bf u}) \propto \mbox{det}\>
\begin{array}{|ccc|}
\cos 3u\> f_n(u)& \cos 3u_{2n}\> f_n(u_{2n})& \cos 3u_{2n+1}\> f_n(u_{2n+1})\\
\sin 3u\> f_n(u)& \sin 3u_{2n}\> f_n(u_{2n})& \sin 3u_{2n+1}\> f_n(u_{2n+1})\\
 g_n(u)& g_n(u_{2n})&  g_n(u_{2n+1})
\end{array}  \nonumber \\
&& \nonumber \\
&& \\
&& f \longleftrightarrow g \nonumber
\end{eqnarray}
The sign "$\propto$" means the equality up to some multiplier which does not depend 
on $u$. The second relation is obtained by
interchanging the letters $f$ and $g$ in the first one.
Recall that we write, for brevity, $f_n(u)$ for $f_n(u,u_1,...u_{2n-1})$,
$f_n(u_{2n})$ for $f_n(u,u_1,...u_{2n-1})$ and so on.
Inserting expressions (\ref{fdef}) and (\ref{gdef}) into equations (\ref{fgrec})
we obtain recurrence relations for the state sum $Z_n(u)$ and its companion $V_n(u)$.

\begin{eqnarray}
 \label{ZVrec}
&&Z_{n+1}({\bf u}) \propto 
\frac{1}{\sin({\bf u}-u_{2n}) \sin({\bf u}-u_{2n+1}) \sin(u_{2n+1}-u_{2n})} \times \nonumber \\
&&\mbox{det}\>\begin{array}{|ccc|}
\cos 3u\> Z_n(u)& \cos 3u_{2n}\> Z_n(u_{2n})& \cos 3u_{2n+1}\> Z_n(u_{2n+1})\\
\sin 3u\> Z_n(u)& \sin 3u_{2n}\> Z_n(u_{2n})& \sin 3u_{2n+1}\> Z_n(u_{2n+1})\\
 V_n(u)& V_n(u_{2n})&  V_n(u_{2n+1})
\end{array} \\
&& \nonumber \\
&& Z \longleftrightarrow V \nonumber
\end{eqnarray}
Recall that the symmetric functions $Z_n(u,u_1,...,u_{2n-1})$ and $V_n(u,u_1,...,u_{2n-1})$ are
 trigonometric polynomials of degree $n-1$ and $n$ correspondingly.
One can easily check that both sides of equation (\ref{ZVrec}) have the same degree in variables 
$\{u,u_{2n},u_{2n+1}\}$. As far as variables $\{u_1,u_2,...,u_{2n-1}\}$ are concerned,
the degree of the right hand side is larger than the degree of the left hand side
by $2n-2$. It is clear that the former has the additional factor depending on variables
$\{u_1,u_2,...,u_{2n-1}\}$. Denote it by $S_n \equiv S_n(u_1,u_2,...,u_{2n-1})$.

The first of relations (\ref{ZVrec}) can be written as
\begin{eqnarray}
\label{Srec}
&&Z_{n+1}({\bf u}) =
\frac{\mbox{det}\> \hat{Z}_n}{S_n\>\sin({\bf u}-u_{2n}) \sin({\bf u}-u_{2n+1})
\sin(u_{2n+1}-u_{2n})} 
\end{eqnarray}
where $\hat{Z}_n$ is the same 3 by 3 matrix as in equation (\ref{ZVrec}).
Let us return for a moment to the variables $\{x\}$ and $\{y\}$ (see equation (\ref{xyunion})).
One can show \cite{KIB} (see also the subsequent papers \cite{Ku1,Ku2,RSasm})
that in the case of $y_{n+1}=x_{n+1}+\eta/2$ 
there exists a simple recurrence relation 
\begin{eqnarray}
 \label{inv}
&&Z_{n+1}(x_1,...,x_n,x_{n+1};y_1,...,y_n,y_{n+1})|_{y_{n+1}=x_{n+1}-\eta/2} = \sin \eta 
\times \\ 
&&\times Z_n(x_1,...,x_n;y_1,...,y_n)
\prod_{i=1}^{n} \sin(\eta-x_{n+1}+x_i)\> \sin(\eta/2+x_{n+1}-y_i). \nonumber
\end{eqnarray}
 In the case $\eta=2\pi/3$ these relations become symmetric in the union 
$\{x\} \cup \{y\}$. Equation (\ref{inv}) can be rewritten as\footnote{we use the symmetry 
of the partition function in variables $\{u\}$}:
\begin{eqnarray}
 \label{invnew}
&&Z_{n+1}(u_0,...,u_{2n+1})|_{u_{2n+1}=u_{2n}-\pi/3} = \frac{\sqrt{3}}{2}
\times \\ 
&&\times Z_n(u_0,...,u_{2n-1})
\prod_{i=0}^{2n-1} \sin(\pi/3+u_{2n}-u_i). \nonumber
\end{eqnarray}
In fact, the same equation is valid for $V_n(\{x\},\{y\})$, as well \cite{Ku2}. 

By using the determinant representations (\ref{IKdet}) and (\ref{Kupdet}) we obtain
the periodicity equations
\begin{eqnarray}
\label{period}
&&Z_{n}(u+\pi)=(-1)^{n-1}\>Z_{n}(u),\quad V_{n}(u+\pi)=(-1)^n\>V_{n}(u).
\end{eqnarray}

Let us put $u_{2n+1}=u_{2n}+2\pi/3$ in equation (\ref{Srec}).
By using equations (\ref{invnew}) and (\ref{period}) we get the expression for the
new sequence of functions $S_n$
\begin{eqnarray}
 \label{S}
&&S_{n}(u_1,...,u_{2n-1}) =(-1)^{n+1} \frac{16}{3}\> 
\biggl(\prod_{i=1}^{2n-1} \sin (\pi/3+a-u_i)\biggr)^{-1}
\times \nonumber \\ 
&&\times \biggl[V_n(a,u_1,...,u_{2n-1})\>Z_n(a+2\pi/3,u_1,...,u_{2n-1})- \\
&&-Z_n(a,u_1,...,u_{2n-1})\>V_n(a+2\pi/3,u_1,...,u_{2n-1})\biggr].\nonumber
\end{eqnarray}
We use notation $a$ rather than $u_{2n}$ to stress the independence of the right hand side
on this variable.
Let us put $a=u_{2n-1}$  into equation (\ref{S}). 
A new expression for the function
$S_n$ can be obtained by using again equations (\ref{invnew}) and (\ref{period}):
\begin{eqnarray}
 \label{Snew}
&&S_{n}(u_1,...,u_{2n-1}) = \frac{16}{3}\> 
\biggl\{V_n(u_{2n-1},u_1,...,u_{2n-1})\>Z_{n-1}(u_1,...,u_{2n-2})+ \nonumber \\
&&+Z_n(u_{2n-1},u_1,...,u_{2n-1})\>V_{n-1}(u_1,...,u_{2n-2})\biggr\}.
\end{eqnarray}

The second of relations (\ref{ZVrec}) leads to the same functions $S_n$ as above. 
Hence we obtain the closed recurrence system. 
It can be used in a different way. In the next section we use this system to find
the refined enumeration of the half-turn symmetric alternating sign matrices.

\begin{center}{\bf4. Application of the recurrence relations}
\end{center}
Let us put $u_1=u_2=...=u_{2n-1}=u_{2n}=0$.
Inserting these values into equation (\ref{Srec}) we obtain
\begin{eqnarray}
&&Z_{n+1}(u,v,0,...,0)=
\frac{1}{S^{(0)}_n \> \sin u \sin(u-v) \sin v} \times \nonumber \\
&&\mbox{det}\>\begin{array}{|ccc|}
\cos 3u\> Z_n(u)&  Z_n& \cos 3v\> Z_n(v)\\
\sin 3u\> Z_n(u)& 0 & \sin 3v\> Z_n(v)\\
 V_n(u)& V_n&  V_n(v)
\end{array}, \nonumber
\end{eqnarray}
where we write $v$ for $u_{2n+1}$.
The right-hand side depends on two functions:
\begin{eqnarray}
&&Z_n(u) \equiv Z_n(u,0,...,0), \quad V_n(u) \equiv V_n(u,0,...,0),\nonumber
\end{eqnarray}
and two constants:
\begin{eqnarray}
&&Z_n \equiv Z_n(0,0,...,0), \quad V \equiv V_n(0,0,...,0), \nonumber
\end{eqnarray}
The constant $S^{(0)}_n \equiv S_n(0,0,...,0)$ is fixed by equation (\ref{Snew})
\begin{eqnarray}
&&S^{(0)}_n=\frac{16}{3} (V_n Z_{n-1}+Z_n V_{n-1}), \nonumber
\end{eqnarray}
In order to close the recurrence relations we put $v=0$ and obtain
\begin{eqnarray}
&&Z_{n+1}(u)=
\frac{3}{16\>(V_n Z_{n-1}+Z_n V_{n-1}) \> \sin^2 u} \times \nonumber \\
&&\mbox{det}\>\begin{array}{|ccc|}
\cos 3u\> Z_n(u)&  Z_n& Z_n^{\prime}(0)\\
\sin 3u\> Z_n(u)& 0 &3 Z_n\\
 V_n(u)& V_n&  V_n^{\prime}(0)
\end{array}. \nonumber
\end{eqnarray}
One can easily check that functions $Z_1(u)$ and $V_1(u)$ are even. 
Using the last equation and its $Z \leftrightarrow V$ companion
one can prove by mathematical induction that functions $Z_n(u)$ and $V_n(u)$ are even
for all natural $n$ and consequently $Z_n^{\prime}(0)=V_n^{\prime}(0)=0$.
The last equation  become simpler
\begin{eqnarray}
\label{Recurrence}
&&Z_{n+1}(u)=
\frac{9\> Z_n \>(Z_n V_n(u)-\cos 3u\> V_n\> Z_n(u))}{16\>(V_n Z_{n-1}+Z_n V_{n-1}) \> \sin^2 u} 
\end{eqnarray}
We consider the case $\eta=2\pi/3$ and all Boltzmann weights defined by 
equation (\ref{weights}) are equal to $\sqrt{3}/2$ if all spectral parameters are equal to $0$.
So we get
\begin{eqnarray}
&&Z_n= (\sqrt{3}/2)^{n^2} A_n, \nonumber
\end{eqnarray}  
where $A_n$ is the number of the alternating sign matrices of order $n$
(details can be found,for example, in the book by Bressoud~\cite{Br}).
The famous formula for $A_n$ was conjectured by Mills, Robbins and Rumsey~\cite{MRR}.
Their conjecture was proven by Zeilberger~\cite{Z0}.
Another proof, using the Izergin-Korepin determinant was given by Kuperberg~\cite{Ku1}. 
We use only the ratio
\begin{eqnarray}
\label{Aratio}
&&\frac{A_{n+1}}{A_n}={3n+1\choose n} / {2n\choose n}.
\end{eqnarray}  
Recall that according to Kuperberg's paper~\cite{Ku2}, 
the product  $Z_n(\{x\},\{y\}) \times V_n(\{x\},\{y\})$ is equal to the partition
function $Z_n^{HT}(\{x\},\{y\})$ of square ice model with special boundary conditions.
This partition function gives the weighted enumeration of 
the half-turn symmetric alternating sign matrices.
The corresponding lattice has $2 n^2$ vertices (see Figure 7 of paper \cite{Ku2})
and we obtain
\begin{eqnarray}
&&Z_n V_n= (\sqrt{3}/2)^{2 n^2} H_{2n}, \nonumber
\end{eqnarray}  
where $H_{2n}$ is the number of the half-turn symmetric
alternating sign matrices of order $2n$.
The nice formula for $H_{2n}$ was conjectured by  Robbins~\cite{Rob}.
It was proven by Kuperberg~\cite{Ku2}.
And again we use only the ratio
\begin{eqnarray}
\label{Hratio}
&&\frac{H_{2n+2}}{H_{2n}}=\frac{4 {3n\choose n}{3n+3\choose n+1}}
{ 3 {2n\choose n}{2n+2\choose n+1}}
\end{eqnarray}  
Let us introduce two normalized sequence of functions 
\begin{eqnarray}  
&&Z_n(u)=Z_n\>z_n(u), \quad V_n(u)=V_n\>v_n(u). \nonumber
\end{eqnarray}  

By inserting these expressions into equation (\ref{Recurrence}) and into its 
$Z \leftrightarrow V$ 
companion, and taking into account ratios (\ref{Aratio}) and  (\ref{Hratio}) we obtain
\begin{eqnarray}  
\label{final}
&&z_{n+1}(u)=\frac{(2n+1)}{3 (3n+1) \sin ^2 u}
\biggl( v_n(u)-\cos 3u \>z_n(u) \biggr), \nonumber  \\
&& \\
&&v_{n+1}(u)=\frac{(2n+1)}{3 (3n+2) \sin ^2 u} 
\biggl( z_n(u)-\cos 3u \>v_n(u) \biggr). \nonumber
\end{eqnarray}  

It is known that the function $z_n(u)$ are related to the number $A_n(r)$ of alternating sign matrices
of order $n$ whose sole '1' of the most left column is at the $r^{th}$ row:
\begin{eqnarray}
\label{zArelation}
&&z_n(u)=A_n^{-1} (\sqrt{3}/2)^{1-n} \sum_{r=1}^n A_n(r)\> a^{r-1}(u)\> b^{n-r}(u),
\end{eqnarray}
where weights $a(u)$ and $b(u)$ are given by equation (\ref{weights}).
The wonderful story of these numbers can be found in the book by Bressoud~\cite{Br}.
Mills, Robbins and Rumsey discovered  a nice formula for
$A_n(r)$~\cite{MRR}.
Their conjecture which is known as the refined ASM conjecture was proven by
Zeilberger~\cite{Z}, who found the state sum $Z(u)$ extending Kuperberg's
method~\cite{Ku1} (see also~\cite{my1}). 
We use only the ratio
\begin{eqnarray}
\label{Arat}
&&\frac{A_n(r)}{A_n}=\frac{(2n-1)! (n+r-2)! (2n-r-1)!}{(n-1)! (3n-2)! (r-1)! (n-r)!}.
\end{eqnarray}  
We find from a comparison of Figure 4  and Figure 7 of the rich in results Kuperberg's
paper~\cite{Ku2} a similar relation
\begin{eqnarray}
\label{zvHrelation}
&&z_n(u) v_n(u)=H_{2n}^{-1} (\sqrt{3}/2)^{1-2n} \sum_{r=1}^{2n} 
H_{2n}(r)\> a^{r-1}(u)\> b^{2n-r}(u),
\end{eqnarray}
where $H_{2n}(r)$ is the number of the half-turn symmetric alternating sign matrices
of order $2n$ whose sole '1' of the most left column is at the $r^{th}$ row.

By using the first of equations (\ref{final}) we can express 
the function $v_n(u)$ in terms of the functions
$z_n(u)$ and $z_{n+1}(u)$
\begin{eqnarray}  
&&v_{n}(u)=\frac{3 (3n+1) (1-\cos 2u)}{2 (2n+1)} z_{n+1}(u) + \cos 3u \>z_n(u). \nonumber
\end{eqnarray}  
Let us multiply this equality by $z_n(u)$. By inserting the right-hand sides of equations
(\ref{zArelation}) and (\ref{zvHrelation}) into the resulting expression we obtain 
\begin{eqnarray}
&&H_{2n}^{-1} \sum_{r=1}^{2n} H_{2n}(r)\> a^{r-1}(u)\> b^{2n-r}(u) = 
\frac{3 (3n+1) (1-\cos 2u)}{2 (2n+1)A_n A_{n+1}}\times \nonumber \\
&& \times \biggl( \sum_{r=1}^n A_n(r)\> a^{r-1}(u)\> b^{n-r}(u) \biggr ) 
\biggl( \sum_{s=1}^{n+1} A_{n+1}(s)\> a^{s-1}(u)\> b^{n-s+1}(u)\biggr)+\nonumber \\ 
&&+\frac{\sqrt{3} \cos 3u}{2 A_n^{2}} \biggl( \sum_{r=1}^n A_n(r)\> a^{r-1}(u)\> b^{n-r}(u)\biggr )
\biggl( \sum_{s=1}^n A_n(s)\> a^{s-1}(u)\> b^{n-s}(u)\biggr ).\nonumber
\end{eqnarray}
After dividing the both sides by $a^{2n-1}$ and changing variables to
\begin{eqnarray}  
&&t \equiv \frac{b(u)}{a(u)}=\frac{\sin (\pi /3-u)}{\sin (\pi /3+u)}, \nonumber
\end{eqnarray} 
we receive the relation between the generating functions ${\cal H}_{2n}(t)$ and ${\cal A}_{n}(t)$
\begin{eqnarray}
\label{Ht}
&&\frac{{\cal H}_{2n}(t)}{H_{2n}} = \frac{{\cal A}_{n}(t)}{ A_{n} (t^2-t+1)} \times \\
&&\times \biggl( \frac{9 (3n+1) (t-1)^2}{4 (2n+1)}\frac{{\cal A}_{n+1}(t)}{A_{n+1}}-
\frac{(t+1)(t-2)(2t-1)}{2} \frac{{\cal A}_n(t)}{A_n}\biggr ),\nonumber
\end{eqnarray}
where 
\begin{eqnarray}
&&{\cal H}_{2n}(t)\equiv \sum_{r=1}^{2n} H_{2n}(r) t^{r-1},\quad 
{\cal A}_{n}(t)\equiv \sum_{r=1}^{n} A_{n}(r) t^{r-1}. \nonumber
\end{eqnarray}

By using equation (\ref{Arat}) one can transform the second multiplier 
in the right-hand side of equation (\ref{Ht})

\begin{eqnarray}
&&\frac{{\cal H}_{2n}(t)}{H_{2n}} = 
\frac{{\cal A}_{n}(t)}{ A_{n}}\>{\cal B}_{n}(t), \nonumber
\end{eqnarray}
where 
\begin{eqnarray}
&&{\cal B}_{n}(t)\equiv \frac{(3n-2) (2n-1)!}{(n-1)!(3n-1)!} \times \nonumber \\
&& \times \sum_{r=1}^{n+1}
\frac{(n^2 -  n r + (r - 1)^2)(n + r - 3)!(2 n - r - 1)!\> 
t^{r-1}}{(r - 1)!(n - r + 1)!} \nonumber
\end{eqnarray}

Therefore the refined enumeration of the half-turn symmetric alternating sign matrices
is given by

\begin{eqnarray}
\label{Result}
&&\frac{H_{2n}(r)}{H_{2n}} = \frac{(2n-1)!(2n-1)!}{(n-1)!(n-1)!(3n-3)!(3n-1)!} \times \nonumber \\
 && \times \sum_{s=1}^{r}\biggl \{
\frac{(n^2 -  n s + (s - 1)^2)(n + s - 3)!(2 n - s - 1)!} 
{(s - 1)!(n - s + 1)!} \times \\
&&\times \frac{(n+r-s-1)!(2n-r+s-2)!} 
{(r-s)!(n-r+s-1)!} \biggr \}, \quad r=1,2,...,n \nonumber
\end{eqnarray}
It is evident that 
\begin{eqnarray}
&&\frac{H_{2n}(r)}{H_{2n}} = \frac{H_{2n}(2n-r+1)}{H_{2n}}, \quad r=n+1,...,2n. \nonumber
\end{eqnarray}

{\it Acknowledgments} I would like to thank A.~V.~Razumov for his valuable comments. 
The work was supported in part by the Russian Foundation for Basic Research under grant
\# 04--01--00352  and by the INTAS under grant \# 00--00561.

\end{document}